\documentclass[prc,twocolumn,superscriptaddress,showpacs,twoside,floatfix,nofootinbib]{revtex4-2}
\usepackage{amssymb,epsfig}
\usepackage{graphicx}
\usepackage{bm}
\usepackage{color}
\usepackage{mathtools}
\usepackage{scrextend}
\emergencystretch=\maxdimen
\usepackage[normalem]{ulem} 
\begin{document}

\title{Observation and spectroscopy of proton-unbound nucleus $^{21}$Al} 
\author{D.~Kostyleva}
\affiliation{GSI Helmholtzzentrum f\"{u}r Schwerionenforschung GmbH, 64291 Darmstadt, Germany}

\author{X.-D.~Xu}
\email{Corresponding author: xiaodong.xu@impcas.ac.cn}
\affiliation{Institute of Modern Physics, Chinese Academy of Science, Lanzhou 730000, China}
\affiliation{GSI Helmholtzzentrum f\"{u}r Schwerionenforschung GmbH, 64291 Darmstadt, Germany}

\author{I.~Mukha}
\affiliation{GSI Helmholtzzentrum f\"{u}r Schwerionenforschung GmbH, 64291 Darmstadt, Germany}

\author{L.~Acosta}
\affiliation{Instituto de Estructura de la Materia, CSIC, 28006, Madrid, Spain}
\affiliation{Instituto de F\'isica, Universidad Nacional Aut\'onoma de M\'exico, M\'exico, Mexico City 01000, Mexico}

\author{M.~Bajzek}
\affiliation{GSI Helmholtzzentrum f\"{u}r Schwerionenforschung GmbH, 64291 Darmstadt, Germany}
\affiliation{II.Physikalisches Institut, Justus-Liebig-Universit\"at, 35392 Gie{\ss}en, Germany}
\affiliation{Faculty of Science, University of Zagreb, 10000 Zagreb, Croatia}

\author{E.~Casarejos}
\affiliation{CINTECX, Universidade de Vigo, DSN, Dpt. Mech. Engineering, E-36310 Vigo, Spain}

\author{A.A.~Ciemny}
\affiliation{Faculty of Physics, University of Warsaw, 02-093 Warszawa, Poland}

\author{D.~Cortina-Gil}
\affiliation{Universidade de Santiago de Compostela, 15782 Santiago de Compostela, Spain}

\author{W.~Dominik}
\affiliation{Faculty of Physics, University of Warsaw, 02-093 Warszawa, Poland}

\author{J.A.~Due\~nas}
\affiliation{Departamento de Ingenieria Electrica y Centro de Estudios Avanzados en Fisica, Matem\'{a}ticas y Computaci\'{o}n, Universidad de Huelva, 21071 Huelva, Spain}

\author{J.~M.~Espino}
\affiliation{Department of Atomic, Molecular and Nuclear Physics, University of Seville, 41012 Seville, Spain}

\author{A.~Estrad\'{e}}
\affiliation{University of Edinburgh, EH1 1HT Edinburgh, United Kingdom}
\affiliation{Central Michigan University, Mt.\ Pleaseant, MI 48859 USA}

\author{F.~Farinon}
\affiliation{GSI Helmholtzzentrum f\"{u}r Schwerionenforschung GmbH, 64291 Darmstadt, Germany}

\author{A.~Fomichev}
\affiliation{Flerov Laboratory of Nuclear Reactions, JINR, 141980 Dubna, Russia}

\author{H.~Geissel}\email{Deceased.}
\affiliation{GSI Helmholtzzentrum f\"{u}r Schwerionenforschung GmbH, 64291 Darmstadt, Germany}
\affiliation{II.Physikalisches Institut, Justus-Liebig-Universit\"at, 35392 Gie{\ss}en, Germany}

\author{J.~G\'{o}mez-Camacho}
\affiliation{Department of Atomic, Molecular and Nuclear Physics, University of Seville, 41012 Seville, Spain}

\author{A.~Gorshkov}
\affiliation{Flerov Laboratory of Nuclear Reactions, JINR, 141980 Dubna, Russia}

\author{L.V.~Grigorenko}
\affiliation{Flerov Laboratory of Nuclear Reactions, JINR, 141980 Dubna, Russia}
\affiliation{National Research Nuclear University ``MEPhI'', 115409 Moscow, Russia}
\affiliation{National Research Centre ``Kurchatov Institute'', Kurchatov square 1, 123182 Moscow, Russia}

\author{Z.~Janas}
 \affiliation{Faculty of Physics, University of Warsaw, 02-093 Warszawa, Poland}

\author{G.~Kami\'{n}ski}
\affiliation{Heavy Ion Laboratory, University of Warsaw, 02-093 Warszawa, Poland}
\affiliation{Flerov Laboratory of Nuclear Reactions, JINR, 141980 Dubna, Russia}

\author{O.~Kiselev}
\affiliation{GSI Helmholtzzentrum f\"{u}r Schwerionenforschung GmbH, 64291 Darmstadt, Germany}

\author{R.~Kn\"{o}bel}
\affiliation{GSI Helmholtzzentrum f\"{u}r Schwerionenforschung GmbH, 64291 Darmstadt, Germany}
\affiliation{II.Physikalisches Institut, Justus-Liebig-Universit\"at, 35392 Gie{\ss}en, Germany}

\author{A.A.~Korsheninnikov}
\affiliation{National Research Centre ``Kurchatov Institute'', Kurchatov square 1, 123182 Moscow, Russia}

\author{S.~Krupko}
\affiliation{Flerov Laboratory of Nuclear Reactions, JINR, 141980 Dubna, Russia}

\author{M.~Kuich}
\affiliation{Faculty of Physics, Warsaw University of Technology, 00-662 Warszawa, Poland}
\affiliation{Faculty of Physics, University of Warsaw, 02-093 Warszawa, Poland}

\author{N.~Kurz}
\affiliation{GSI Helmholtzzentrum f\"{u}r Schwerionenforschung GmbH, 64291 Darmstadt, Germany}

\author{Yu.A.~Litvinov}
\affiliation{GSI Helmholtzzentrum f\"{u}r Schwerionenforschung GmbH, 64291 Darmstadt, Germany}

\author{G.~Marquinez-Dur\'{a}n}
\affiliation{Department of Applied Physics, University of Huelva, 21071 Huelva, Spain}

\author{I.~Martel}
\affiliation{University of Huelva, 21007 Huelva, Spain}

\author{C.~Mazzocchi}
\affiliation{Faculty of Physics, University of Warsaw, 02-093 Warszawa, Poland}

\author{E.Yu.~Nikolskii}
\affiliation{Flerov Laboratory of Nuclear Reactions, JINR, 141980 Dubna, Russia}
\affiliation{National Research Centre ``Kurchatov Institute'', Kurchatov square 1, 123182 Moscow, Russia}
	
\author{C.~Nociforo}
\affiliation{GSI Helmholtzzentrum f\"{u}r Schwerionenforschung GmbH, 64291 Darmstadt, Germany}

\author{A.K.~Ord\'{u}z}
\affiliation{Grand Acc\'{e}l\'{e}rateur Nacional d'Ions Lourds -- GANIL, 14076 Caen, France}

\author{M.~Pf\"{u}tzner}
\affiliation{Faculty of Physics, University of Warsaw, 02-093 Warszawa, Poland}
\affiliation{GSI Helmholtzzentrum f\"{u}r Schwerionenforschung GmbH, 64291 Darmstadt, Germany}

\author{S.~Pietri}
\affiliation{GSI Helmholtzzentrum f\"{u}r Schwerionenforschung GmbH, 64291 Darmstadt, Germany}

\author{M.~Pomorski}
\affiliation{Faculty of Physics, University of Warsaw, 02-093 Warszawa, Poland}

\author{A.~Prochazka}
\affiliation{GSI Helmholtzzentrum f\"{u}r Schwerionenforschung GmbH, 64291 Darmstadt, Germany}

\author{C.~Rodr\'{i}guez-Tajes}
\affiliation{Universidade de Santiago de Compostela, 15782 Santiago de Compostela, Spain}

\author{S.~Rymzhanova}
\affiliation{Flerov Laboratory of Nuclear Reactions, JINR, 141980 Dubna, Russia}

\author{A.M.~S\'{a}nchez-Ben\'{i}tez}
\affiliation{Centro de Estudios Avanzados en F\'{i}sica, Matem\'{a}ticas y Computaci\'{o}n (CEAFMC), Department of Integrated Sciences, University of Huelva, 21071 Huelva, Spain}

\author{C.~Scheidenberger}
\affiliation{GSI Helmholtzzentrum f\"{u}r Schwerionenforschung GmbH, 64291 Darmstadt, Germany}
\affiliation{II.Physikalisches Institut, Justus-Liebig-Universit\"at, 35392 Gie{\ss}en, Germany}
\affiliation{Helmholtz Research Academy Hesse for FAIR (HFHF), GSI Helmholtz Center for Heavy Ion Research, Campus Gießen, Gießen 35392, Germany}

\author{H.~Simon}
\affiliation{GSI Helmholtzzentrum f\"{u}r Schwerionenforschung GmbH, 64291 Darmstadt, Germany}

\author{B.~Sitar}
\affiliation{Faculty of Mathematics and Physics, Comenius University, 84248 Bratislava, Slovakia}

\author{R.~Slepnev}
\affiliation{Flerov Laboratory of Nuclear Reactions, JINR, 141980 Dubna, Russia}

\author{M.~Stanoiu}
\affiliation{IFIN-HH, Post Office Box MG-6, Bucharest, Romania}

\author{P.~Strmen}\email{Deceased.}
\affiliation{Faculty of Mathematics and Physics, Comenius University, 84248 Bratislava, Slovakia}

\author{K.~S\"{u}mmerer}
\affiliation{GSI Helmholtzzentrum f\"{u}r Schwerionenforschung GmbH, 64291 Darmstadt, Germany}

\author{I.~Szarka}
\affiliation{Faculty of Mathematics and Physics, Comenius University, 84248 Bratislava, Slovakia}

\author{M.~Takechi}
\affiliation{GSI Helmholtzzentrum f\"{u}r Schwerionenforschung GmbH, 64291 Darmstadt, Germany}

\author{Y.K.~Tanaka}
\affiliation{University of Tokyo, 113-0033 Tokyo, Japan}
\affiliation{GSI Helmholtzzentrum f\"{u}r Schwerionenforschung GmbH, 64291 Darmstadt, Germany}

\author{H.~Weick}
\affiliation{GSI Helmholtzzentrum f\"{u}r Schwerionenforschung GmbH, 64291 Darmstadt, Germany}

\author{J.S.~Winfield}\email{Deceased.}
\affiliation{GSI Helmholtzzentrum f\"{u}r Schwerionenforschung GmbH, 64291 Darmstadt, Germany}

\author{P.J.~Woods}
\affiliation{University of Edinburgh, EH1 1HT Edinburgh, United Kingdom}

\author{M.V.~Zhukov}
\affiliation{Department of Physics, Chalmers University of Technology, S-41296 G\"oteborg, Sweden}

\date{\today}

\begin{abstract}

We report on the observation of previously-unknown isotope $^{21}$Al, the first unbound aluminum isotope located beyond the proton dripline. The $^{21}$Al nucleus decays by one-proton (1\textit{p}) emission, and its in-flight decays were detected by tracking trajectories of all decay products with micro-strip silicon detectors. The 1\textit{p}-emission processes were studied by analyses of the measured angular correlations of decay products $^{20}$Mg+\textit{p}. The 1\textit{p}-decay energies of ground and low-lying excited states of $^{21}$Al, its mass excess and proton separation energy value $S_p$=$-1.1(1)$ MeV were determined.

\end{abstract}

\maketitle

{\textbf{Motivation.}}
Nuclear structure and decays beyond the proton dripline were addressed in a number of experimental and theoretical studies of light isotopes, see e.g.~the recent review in Ref.~\cite{Pfutzner:2023} and references therein.~For example in Ref.~\cite{Grigorenko:2018}, the isotopes $^{26}$Ar and $^{25}$Cl were predicted as the most remote nuclear configurations with identified ground states (g.s.)~in their respective isotopic chains.~The predictions used known 1\textit{p} and 2\textit{p} decay mechanisms, which may be applied to a number of unknown unbound nuclei located within a relatively broad (by 2--5 atomic mass units) area along the proton dripline.~The most remote nuclear systems are expected to have no individual resonances, hence, they can not be identified as isotopes anymore.~Therefore, a new borderline indicating the limits of existence of nuclei in the nuclear chart and the transition to chaotic-nucleon matter may be discussed~\cite{Grigorenko:2018,Kostyleva:2019}.

In this work, we continue the ``excursion beyond the proton dripline'' of Ref.~\cite{Mukha:2018} by presenting the results of additional analysis of the data obtained as byproducts of the S271 experiment~\cite{Mukha:2010,Mukha:2012} and S388 experiment~\cite{Xu:2018,Mukha:2018} at the SIS facility at GSI, Germany. The primary aims of these two experiments were investigations of 2$p$ decays in-flight of $^{19}$Mg and $^{30}$Ar isotopes, respectively.~The present study focuses on the previously-unobserved isotope $^{21}$Al and reports the analysis results from both experiments.

{\textbf{Experiment.}}
The experiment S271 was described in detail in Ref.~\cite{Mukha:2010,Mukha:2012}.~Here, a brief description of this experiment and the detector performance are provided. During this experiment, the $^{20}$Mg secondary beam was produced in fragmentation of a 591 $A$MeV $^{24}$Mg primary beam.~The fragment separator (FRS) was operated in a separator-spectrometer mode, where the first half of the FRS was tuned for separation and focusing of a 450 $A$MeV radioactive beam of $^{20}$Mg on a 2 $\rm{g/cm}^2$ $^9$Be secondary target in the middle focal plane (F2) of FRS, and the second half of FRS was set for the detection of heavy-ion (HI) decay products, e.g.~$^{17}$Ne. The secondary $^{20}$Mg beam had a strong admixture of other ions with mass-to-charge ratio similar to $^{20}$Mg. In particular, $^{21}$Mg and $^{22}$Al ions were transported with considerable intensities.~Then $^{21}$Al nuclei could be produced via one-neutron removal, charge-exchange and one-proton pickup of the $^{22}$Al, $^{21}$Mg, $^{20}$Mg projectiles, respectively. The decay products of unbound $^{21}$Al nuclei were tracked by using an array of micro-strip double-sided silicon detectors (DSSD) located just downstream of the secondary target. The array consisted of four large-area DSSDs~\cite{Stanoiu:2008}, which were employed to measure hit coordinates of the protons and the HI decay products, resulting from the in-flight decays of the studied 2$p$ precursors. The high-precision position measurements with DSSDs served for reconstruction of all fragment trajectories, which allowed for deriving the angular HI-$p$ and HI-$p$-$p$ correlations. In the present study of $^{21}$Al, the data obtained from the experiment S388 using the $^{36}$Ar primary beam and similar setup, were also analyzed.~During the experiment S388, the $^{20,21}$Mg and $^{22}$Al ions were produced in fragmentation of a primary 685 \textit{A}MeV $^{36}$Ar beam. In particular, the 460 \textit{A}MeV $^{20}$Mg beam bombarded a 5 g/cm$ ^{2} $ thick $^9$Be secondary target located at the F2 of FRS. A detailed description of the experiment S388 can be found in Refs.~\cite{Mukha:2018,Xu:2018}.

\begin{figure}[!htbp]
\centerline{\includegraphics[width=0.48\textwidth]{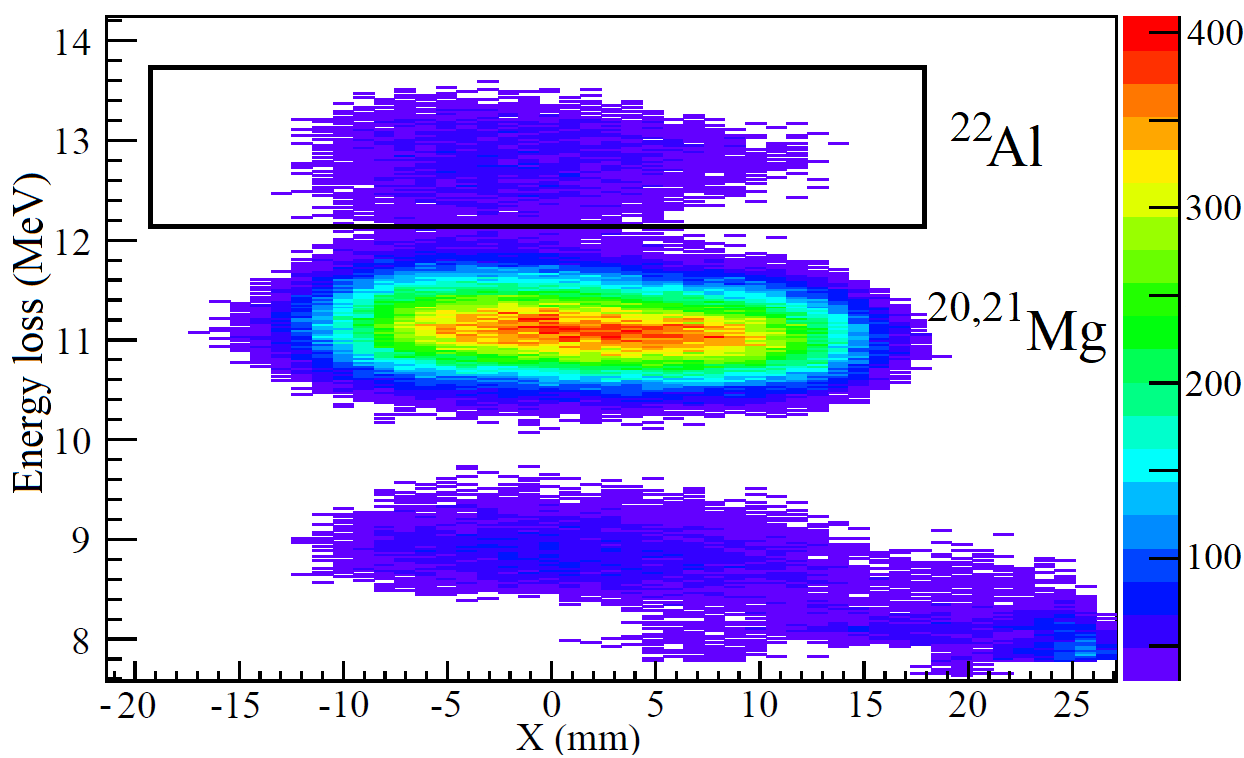}}
\caption{Energy loss of secondary-beam ions in the DSSD located in front of the secondary $^{9}$Be target as a function of their hit coordinate X (which is the transverse coordinate relative to the beam direction) during the experiment S271~\cite{Mukha:2010,Mukha:2012}. Ion intensities are shown according to the scale on right-hand side. Locations of ions of interest $^{22}$Al, $^{20,21}$Mg are marked. The gate for selection of $^{22}$Al projectiles is shown by a black-line rectangle.}
\label{fig:dE_X}
\end{figure}

The described $^{20}$Mg settings of the S388 experiment reproduced those of the previous S271 experiment. There was however a small but important difference between the detector setups. In the S271 experiment, an additional position-sensitive DSSD detector was placed in front of the secondary target.~It measured the energy losses and hit coordinates of all secondary-beam projectiles.~A two-dimensional plot of these observables is presented in Fig.~\ref{fig:dE_X}.~One may see that energy losses of Al and Mg projectiles can be discriminated, in particular events corresponding to ions $^{22}$Al can be selected.

{\textbf{Data.}}
In the experiment S271, the heavy-ion decay products of secondary reactions were unambiguously identified by means of their magnetic rigidity, time of flight, and energy loss.~Corresponding identification plot can be found in Fig.~3 of Ref.~\cite{Mukha:2010}. For example, the $^{21}$Al spectrum was obtained on the basis of the angular correlations of decay products $^{20}$Mg+$p$ by using their measured trajectories. With the purpose of a unambiguous interpretation, the coincident $^{20}$Mg+$p$ events were gated by choosing the Al projectiles as shown in Fig.~\ref{fig:dE_X}, thereby suggesting a population of $^{21}$Al states induced by  neutron knockout reactions. In extracting the $^{21}$Al spectroscopic information, we applied a similar data analysis procedure which was previously used for the 1\textit{p}-decays of $^{15}$F, $^{18}$Na, and $^{29,30}$Cl~\cite{Mukha:2010, Mukha:2012, Mukha:2018}. Those known corresponding calibration coefficients were utilized in this analysis.

\begin{figure}[!htbp]
\centerline{\includegraphics[width=0.49\textwidth]{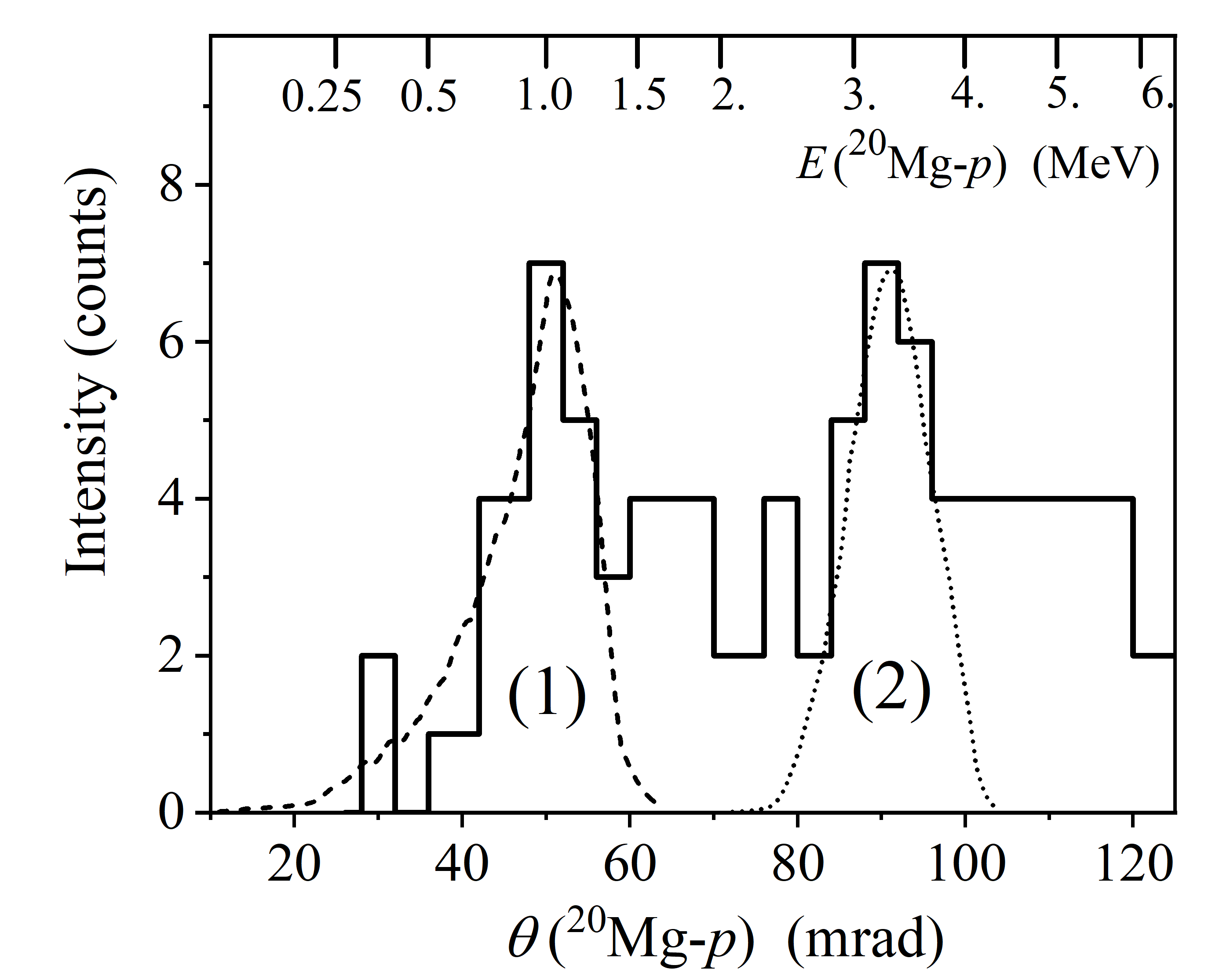}}
\caption{Angular $\theta$(${^{20}\mathrm{Mg-}p}$) correlations derived from the measured $^{20}$Mg+$p$ coincidences obtained in the experiment S271 (histogram). The $^{22}$Al projectiles before the secondary target were selected. The respective $1p$-decay energies $E$(${^{20}\mathrm{Mg-}p}$) are given in the upper axes. The peaks (1) and (2) were reproduced by applying Monte-Carlo simulations of 1\emph{p}-decays of two $^{21}$Al states into the $^{20}$Mg g.s. (shown by the dashed and dotted curves) with the evaluated \emph{1p}-decay energies of 1.1(1) and 3.20(15) MeV, respectively.}
\label{fig:theta-p-Mg_b}
\end{figure}

In Fig.~\ref{fig:theta-p-Mg_b}, we present the angular correlations between $^{21}$Al decay products, namely $^{20}$Mg and proton measured in the experiment S271.~The events in Fig.~\ref{fig:theta-p-Mg_b} were obtained by selecting only $^{22}$Al projectiles at F2 (as illustrated in Fig.~2 of Ref.~\cite{Mukha:2010} and Fig.~\ref{fig:dE_X}). The respective $1p$-decay energies $E$(${^{20}\mathrm{Mg-}p}$) are given in the upper axes. One can see two peaks, which are labeled as (1) and (2) in Fig.~\ref{fig:theta-p-Mg_b}. They indicate the population and 1\textit{p}-decays of two lowest states in $^{21}$Al.

\begin{figure}[!htbp]
\centerline{\includegraphics[width=0.48\textwidth]{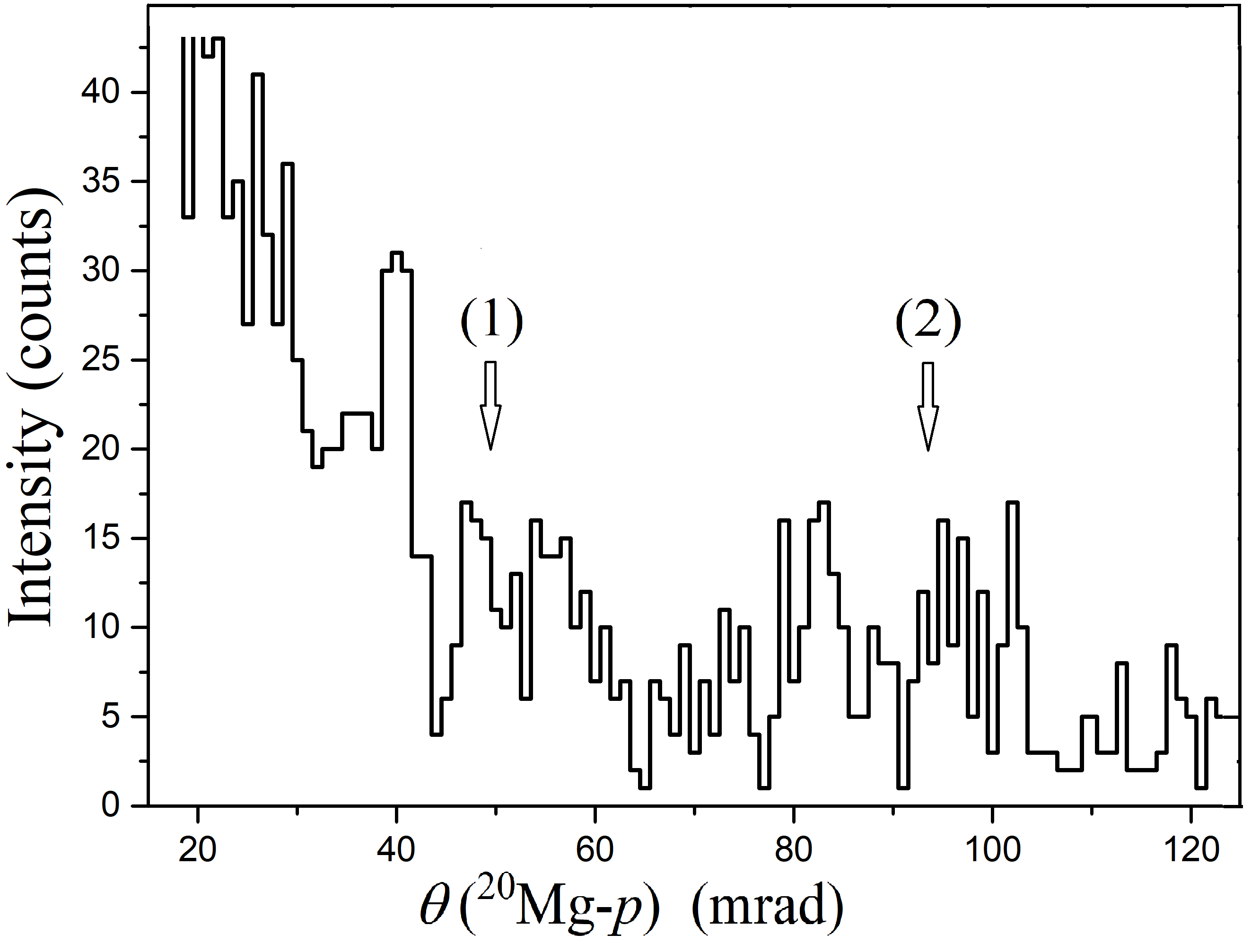}}
\caption{Angular $\theta$(${^{20}\mathrm{Mg-}p}$) correlations derived from the $^{20}$Mg+$p$ coincidences measured in the experiment S271. The contributions of all secondary-beam projectiles (e.g., $^{22}$Al, $^{20,21}$Mg) are included. The arrows point to the positions of peaks induced by $^{22}$Al (i.e., peaks (1) and (2) shown in Fig.\ref{fig:theta-p-Mg_b}).}
\label{fig:theta-p-Mg_b1}
\end{figure}
In order to determine the decay energies of peaks exhibited in Fig.~\ref{fig:theta-p-Mg_b}, we performed Monte Carlo simulations of the detector response to the 1\textit{p}-decays of $^{21}$Al states by using the GEANT software~\cite{Geant_ref}, which was described in detail in Refs.~\cite{Mukha:2010,Mukha:2012}. The simulations were performed independently for the peaks (1) and (2). $Q_{1p}$ values and their uncertainties were derived from the centroids and widths of the distributions of probabilities calculated by using the standard Kolmogorov test, which computes the probability that the simulated $\theta$(${^{20}\mathrm{Mg-}p}$) spectrum matches the respective experimental pattern~\cite{KS_test}. Consequently, the $Q_{1p}$ values for the peak (1) and (2) in Fig.~\ref{fig:theta-p-Mg_b} were determined to be 1.1(1) and 3.20(15) MeV, respectively.

In Fig.~\ref{fig:theta-p-Mg_b1}, we present in addition the ``inclusive'' angular correlations between $^{20}$Mg and proton measured in the experiment S271. The contributions of all secondary-beam projectile ions (e.g., $^{22}$Al, $^{20,21}$Mg) are included. It is worth mentioning that there is an uncertainty associated with the derived $Q_{1p}$ values due to three possible reactions populating the $^{21}$Al states with different intensities: (a) one-neutron removal with $^{22}$Al projectiles, (b) charge-exchange with $^{21}$Mg beam, and (c) proton pickup by $^{21}$Mg projectiles. In the case of unresolved states, the corresponding overlapped peaks may have shifted center-of-gravity positions depending on their population intensities.

The respective $1p$-decay energies $E$(${^{20}\mathrm{Mg-}p}$) can not be assigned unambiguously in Fig.~\ref{fig:theta-p-Mg_b1}, as different projectiles have different momenta. One can see few peaks in addition to the $^{22}$Al-induced peaks (1) and (2) shown in Fig.~\ref{fig:theta-p-Mg_b}. They may indicate the population and 1\textit{p}-decays of the lowest states in $^{21}$Al induced by reactions with $^{20}$Mg or $^{21}$Mg projectiles. In particular, the peak around 55 mrad may reflect either the g.s.~or predicted first-excited state in $^{21}$Al produced by $^{20}$Mg or $^{21}$Mg, respectively. We can not distinguish these two scenarios in our measurement. One may also see a broad distribution in Fig.~\ref{fig:theta-p-Mg_b1} at small $\theta_{^{20}\mathrm{Mg-}p}$ values which should correspond to very low $Q_{1p}$ energies. Its origin is not clear, and a cross-check is required. There is also the lowest peak at 40 mrad, which may correspond to $E$(${^{20}\mathrm{Mg-}p}$)$\simeq$0.6 MeV if it is populated due to the reactions with $^{20}$Mg projectiles. Thus the 40-mrad peak must be examined as a possible candidate for $^{21}$Al ground state.

\begin{figure}[!htbp]
\centerline{\includegraphics[width=0.49\textwidth]{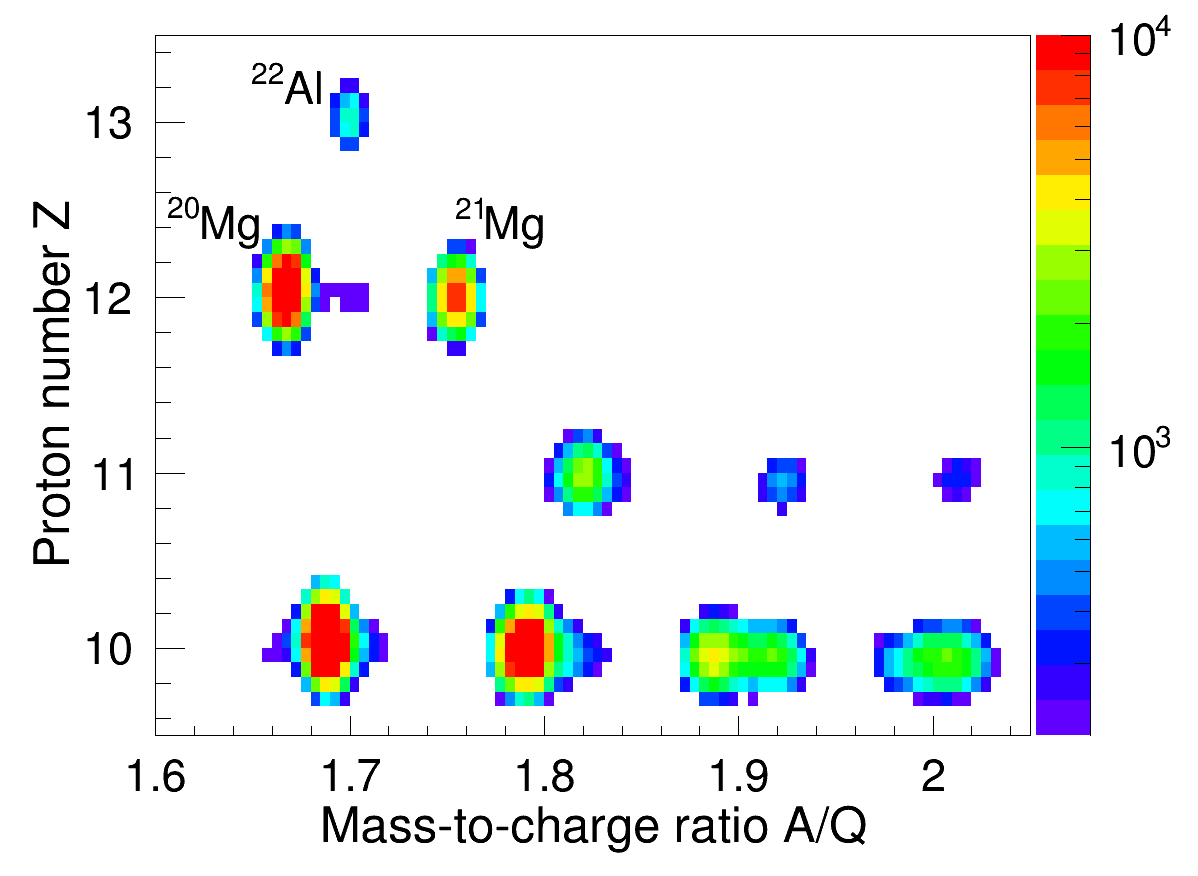}}
\caption{Two-dimensional identification plot of \textit{ Z} vs \textit{A/Q} for the heavy ions detected at FRS during the experiment S388 with the $^{20}$Mg$-^{17}$Ne setting. Projectiles of interest $^{21}$Mg and $^{22}$Al were also produced and transported as by-products with reasonable intensities. Ion intensities are shown according to the logarithmic scale on right-hand side.}
\label{fig:21Al-ID}
\end{figure}

In order to cross-check the states observed in the experiment S271, we analyzed the data obtained during one setting of the experiment S388, in which reactions with a $^{20}$Mg secondary beam producing the known $^{19}$Mg were employed for reference purposes~\cite{Xu:2018}. In this measurement, the first half of the FRS was optimized to transport the $^{20}$Mg beam and the second half of the FRS was tuned to transmit the $^{17}$Ne ions. Fig.~\ref{fig:21Al-ID} shows the particle identification plot for the ions which reached the last focal plane of FRS (F4), where the ions' proton number \textit{Z} versus their mass-to-charge ratio \textit{A/Q} (\textit{Q=Z} in our case of light ions and high energy) are shown. One may see that $^{20}$Mg ions were also transported down to F4, because their \textit{A/Q} ratio similar to that of $^{17}$Ne. Other ions of interest, such as $^{21}$Mg and $^{22}$Al, were also produced and transported through FRS with the signal-to-background ratio larger than 100.

\begin{figure}[!htbp]
\centerline{\includegraphics[width=0.48\textwidth]{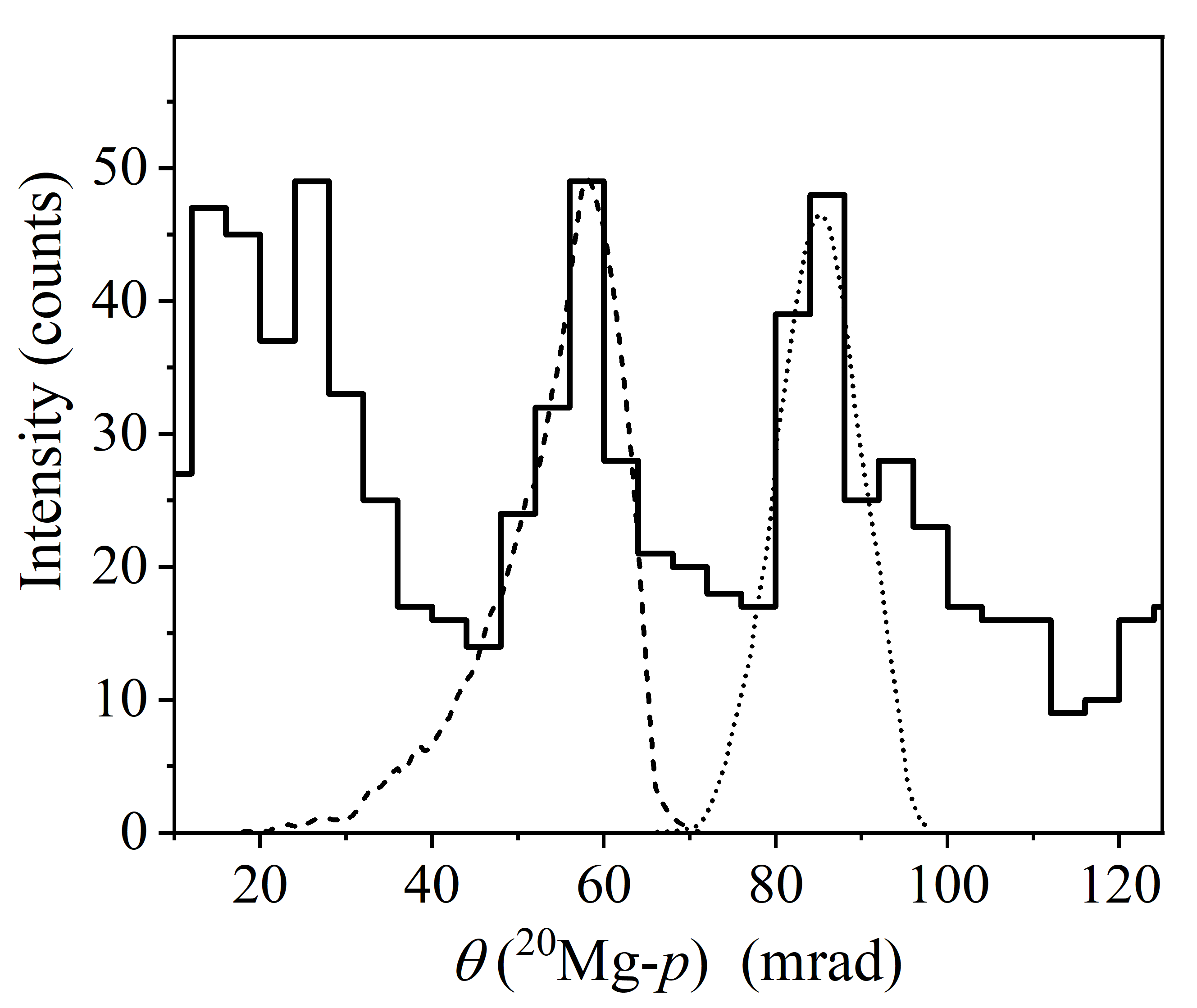}}
\caption{Angular $\theta$(${^{20}\mathrm{Mg-}p}$) correlations derived from the measured $^{20}$Mg+$p$ coincidences obtained in the experiment S388 (histogram). The contributions of all secondary-beam ions illustrated in Fig.~\ref{fig:21Al-ID} are included. The dashed and dotted curves representing the Monte-Carlo simulations of 1\emph{p}-decays of two $^{21}$Al states into the $^{20}$Mg g.s.~are displayed for illustration purposes.}
\label{fig:theta-p-Mg_a}
\end{figure}

Fig.~\ref{fig:theta-p-Mg_a} displays the $\theta$(${^{20}\mathrm{Mg-}p}$) spectra obtained from the $^{20}\rm{Mg}+p$ coincidences measured in the S388 experiment. Two peaks are also present which suggests the similar 1\textit{p}-decays of low-energy states in $^{21}$Al. Since there was no selection of incoming ions before secondary target, the $^{21}$Al isotope may be produced in reactions of one-neutron knockout, charge-exchange and one-proton pickup of the $^{22}$Al, $^{21}$Mg and $^{20}$Mg projectiles, respectively. In comparison with the S271 data shown in Fig.~\ref{fig:theta-p-Mg_b1}, the distribution exhibits a similar shape characterized by two peaks centered around 60 and 90 mrad, accompanied by a broad distribution at smaller $\theta_{^{20}\mathrm{Mg-}p}$ values. The peaks repeat the corresponding structures in Fig.~\ref{fig:theta-p-Mg_b1} though the angular resolution is much worse, and one may assume that the two double-peak structures in Fig.~\ref{fig:theta-p-Mg_b1} merge into the 60- and 90-mrad peaks. For illustration purposes, the simulations similar to those in Fig.~\ref{fig:theta-p-Mg_b} are added, though the corresponding \emph{1p}-decay energies can not be derived accurately because of the ambiguous momenta of projectiles at F2 in this case.

The comparison of the results of Monte-Carlo simulations of two experiments demonstrates a reasonable agreement between the simulated 90-mrad peaks in Fig.~\ref{fig:theta-p-Mg_a} and in Fig.~\ref{fig:theta-p-Mg_b} with the derived $Q_{1p}$ value of 3.20(15) MeV. Both peaks have full width at half maximum (FWHM) values of $\simeq$1.0 MeV. There is also qualitative agreement with an insignificant deviation in the positions of the 50-mrad peaks. The most important difference is that the narrow 40-mrad peak present in Fig.~\ref{fig:theta-p-Mg_b1} is not observed in Fig.~\ref{fig:theta-p-Mg_a}, which eliminates the possible assignment of the 40-mrad peak to the g.s.\ of $^{21}$Al. The data of the S271 experiment have larger statistics, and their interpretation is straightforward because of the dominating secondary reaction $^{22}$Al$\rightarrow ^{21}$Al+\textit{n}. Therefore, we ignore the lowest 40-mrad peak in Fig.~\ref{fig:theta-p-Mg_b1} and assign the $^{21}$Al g.s.~to the 50-mrad peak in Fig.~\ref{fig:theta-p-Mg_b} with the derived 1\emph{p}-decay energy of 1.1(1) MeV. The FWHM value of the peak (1) in Fig.~\ref{fig:theta-p-Mg_b} is $\simeq$0.5 MeV. The corresponding estimate of the $^{21}$Al g.s.~width derived by fitting these peaks provides only the upper-limit value $\Gamma_{g.s.}<$400 keV, which is mainly due to the experimental resolution.

\begin{figure}[!htbp]
\centerline{\includegraphics[width=0.42\textwidth]{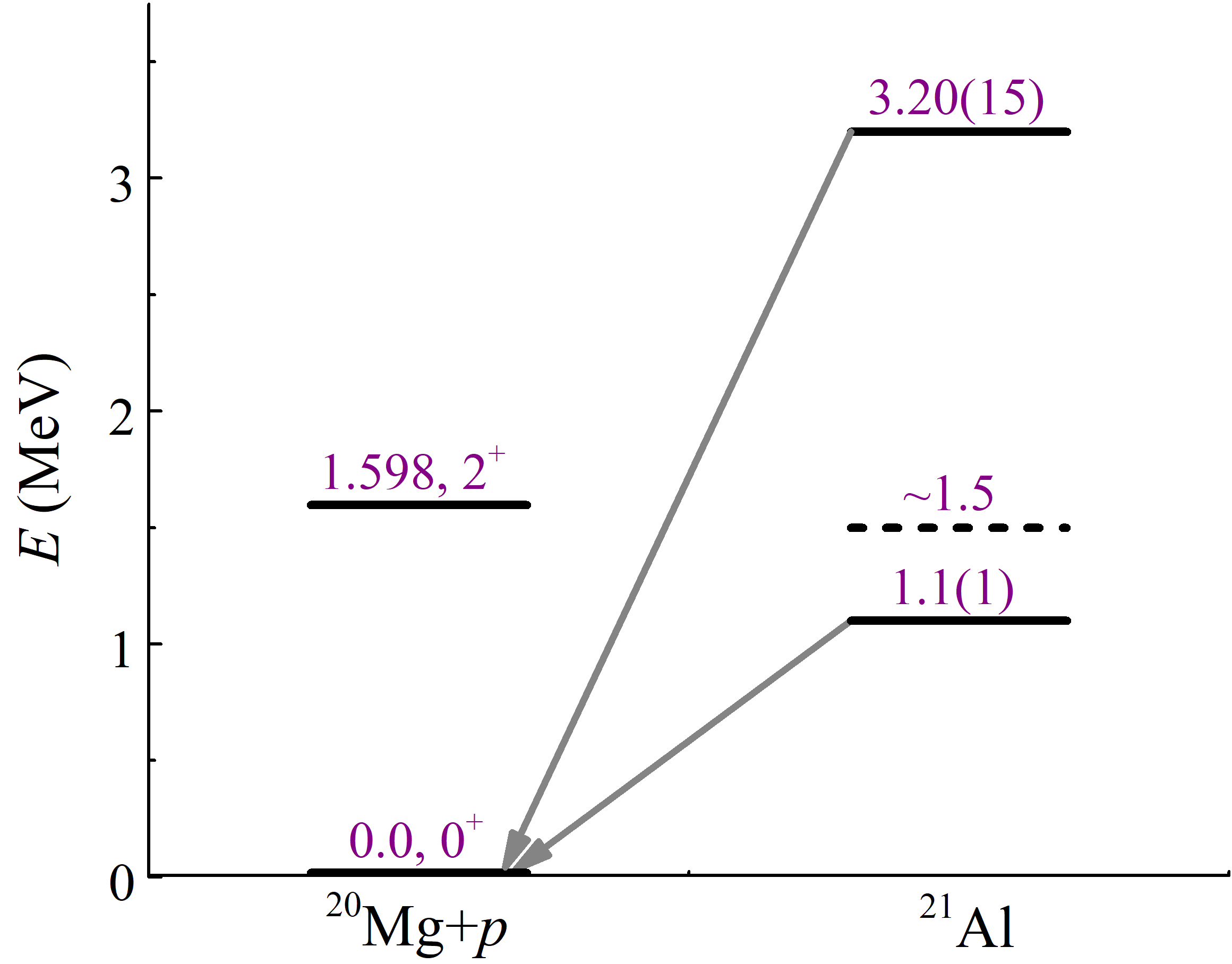}}
\caption{Proposed decay scheme of two low-energy states observed in $^{21}$Al with tentatively-assigned \emph{1p}-decay channels through the known $^{20}$Mg ground state, whose energies are given relative to the 1$p$ threshold. The dashed line at 1.5 MeV shows the tentative position of the first excited state indicated in data in Fig.~\ref{fig:theta-p-Mg_b1}.}
\label{fig:decay-scheme}
\end{figure}

{\textbf{Discussion.}} 
The assigned levels and decay scheme of $^{21}$Al derived from the observed angular correlations is shown in Fig.~\ref{fig:decay-scheme}.~The g.s.~of $^{21}$Al is unbound with 1\textit{p}-decay energy of 1.1(1) MeV. The mass of $^{21}$Al can be derived by using the masses of $^{20}$Mg and proton together with the $Q_{1p}$ value, which then can be compared with available theoretical predictions.~The AME2020 atomic mass evaluation~\cite{Kondev:2021} predicts that mass excess (M.E.) of the $^{21}$Al g.s.~is 27.1(6) MeV, which exceeds the measured M.E.~value of 25.9(1) MeV by 1.2 MeV. Such a difference with AME2020 may be explained due to the effect of Thomas-Ehrman shift~\cite{Thomas:1952,Ehrman:1951} which is often observed in \emph{1p}-unbound nuclei. These level shifts may be described by the empirical $S_p$ systematics of \emph{1p}-emitting $d_{5/2}$ states in light nuclei. The systematics is based on a parameterization of a mirror energy difference MED~\cite{Fortune:2018} by using the available experimental data on neutron-rich nuclei.~The MED definition is MED=$S_n-S_p$ (here $S_n$, $S_p$ are 1\textit{n} and 1\textit{p} separation energies in mirror states of neutron- and proton- rich nuclei, respectively), and its parametrization is MED=(\emph{Z/A}$^{1/3}$)MED', where the MED' value does not depend on the proton number \emph{ Z} and mass number \emph{A}~\cite{Fortune:2018}. This parameterization predicts $S_{p}$=-1.315 MeV for $^{21}$Al (by using the $S_{n}$ value of its mirror partner $^{21}$O$_{g.s.}$(5/2$^{+}$) and the corresponding MED value~\cite{Fortune:2018}), which is in reasonable agreement with the measured $S_{p}$ value of -1.1(1) MeV. The mass of $^{21}$Al g.s.~was also predicted by the systematics proposed for the mass differences of mirror nuclei (the improved Kelson-Garvey mass relations~\cite{Tian:2013}). The estimated value $S_{p}$=-1.265(13) MeV agrees with the data.

There are several theoretical predictions of the low-energy states and their widths dedicated to $^{21}$Al, e.g.~see Refs.~\cite{Timofeyuk:2012,Holt:2013,Michel:2019,Ma:2021,Li:2023}.~Let us consider the predictions calculated by using both the microscopic cluster model and potential model~\cite{Timofeyuk:2012}. The mirror symmetry between the widths of low-lying resonance states and the asymptotic normalization coefficients of their mirror analogs were employed. The potential model used the same potential well parameters obtained from descriptions of the respective mirror states in $^{21}$O which are known experimentally.~The predicted 1\textit{p}-decay energies and widths ($E_1p$, $\Gamma$) of the lowest $^{21}$Al states with $d_{5/2}$, $s_{1/2}$, and $d_{3/2}$ 1\textit{p}-configurations are (1.21 MeV, 0.7 keV), (1.55 MeV, 248 keV), and (2.63 MeV, $\lessapprox$0.08 keV), respectively~\cite{Timofeyuk:2012}. The calculated $d_{5/2}$-state energy is close to the measured value of the $^{21}$Al g.s., but the corresponding width is too small to be obtained with our setup, which is able however to measure the predicted width of the first-excited $s_{1/2}$ state in a future dedicated experiment. One should note that states in $^{21}$Al with core $^{20}$Mg$^{*}$(2$^{+}$) excitations (predicted by the microscopic cluster model in Ref.~\cite{Timofeyuk:2012}) were not identified in the present work as their identification requires an additional detection of $\gamma$ rays from the fragments $^{20}$Mg$^{*}$.

\begin{figure}[!htbp]
\centerline{\includegraphics[width=0.48\textwidth]{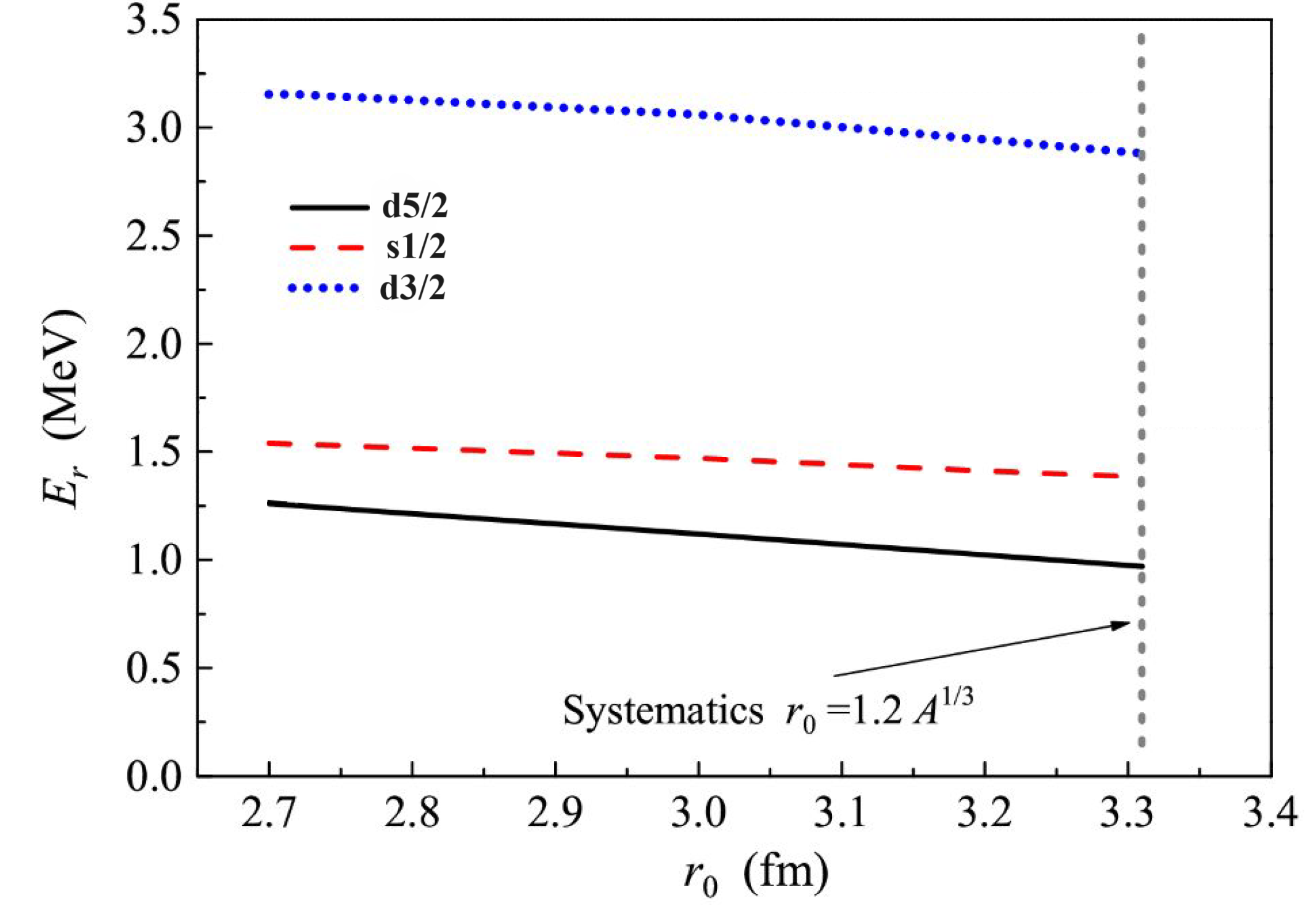}}
\caption{Energies $E_r$ of the lowest states in $^{21}$Al with $s_{1/2}$, $d_{5/2}$ and $d_{3/2}$ 1\textit{p} configurations (the dashed, solid and dotted curves, respectively) as functions of nuclear radius $r_0$.~The $E_r$ values were calculated relative to the $^{20}$Mg+\textit{p} threshold by applying a potential model~\cite{Grigorenko:2018}. The vertical dashed line shows the radius obtained from the expression $r_0$=1.2$\cdot A^{1/3}$.}
\label{fig:energies}
\end{figure}

The radial dependence of energies of $^{21}$Al low-lying states with $d_{5/2}$, $s_{1/2}$, and $d_{3/2}$ 1\textit{p}-configurations were also calculated by using another nuclear potential model~\cite{Grigorenko:2018} in a manner similar to Ref.~\cite{Timofeyuk:2012}.~The calculated level energies are illustrated in Fig.~\ref{fig:energies} as functions of nuclear radius, where the trend of decreasing level energies as the radius increases demonstrates the effect of decreased Coulomb repulsion.~One may see that the energies of states with 5/2$^{+}$ and 3/2$^{+}$ configurations match the measured energies of 1.1 and 3.2 MeV derived from the peaks in Fig.~\ref{fig:theta-p-Mg_b}, respectively.~According to the predictions shown in Fig.~\ref{fig:energies}, the energy of the 1/2$^{+}$ configuration is expected only about 0.4 MeV higher than the g.s.~of $^{21}$Al. It is worth mentioning that the excitation energy of the first excited state (1/2$^{+}$) in $^{21}$Al was also calculated within the many-body perturbation theory~\cite{Holt:2013} and the Gamow shell-model~\cite{Michel:2019}, which is in agreement with the value provided by the applied potential model.~Such a relatively small energy difference between the ground state and the first excited state indicate that the low-energy peaks (within the range of 45--65 mrad) in Fig.~\ref{fig:theta-p-Mg_b1} may consist of the 5/2$^{+}$ and 1/2$^{+}$ states which were unresolved in the S388 measurement due to the limited energy resolution. In Fig.~\ref{fig:theta-p-Mg_b}, there is an indication on possible population of the first excited state at around 1.5 MeV, for which the un-explained part of the angular $\theta_{^{20}\mathrm{Mg}-p}$ distribution points at 65 mrad.

It is also worth noting that the spectrum of $^{21}$O, which is the mirror nucleus of $^{21}$Al, exhibits several low-energy levels with excitation energy below the one-neutron threshold of 3.8 MeV. Considering the mirror symmetry, more low-energy states in $^{21}$Al may be found. For example, an indication on possible excitations in $^{21}$Al can be spotted in Fig.~\ref{fig:theta-p-Mg_b} in the region of 4--6 MeV (i.e., within the range of 100--120 mrad). Though the mentioned regions don't have sufficient statistics for a quantitative investigation, future experiments with higher statistics and better resolution will resolve these questions.

We estimated the half-lives of the observed $^{21}$Al states by measuring distributions of their decay vertexes in the same way as in the previous study of 2\textit{p} decay of $^{30}$Ar~\cite{Mukha:2015}. All vertex distributions are located within the reaction target, and therefore we found no indication on long-lived states in $^{21}$Al, which results in assignment of the upper limit of half-life of 10 ps.

{\textbf{Conclusion.}} 
The first spectroscopy of the previously-unknown isotope $^{21}$Al which decays via 1\textit{p} emission, revealed two low-energy states which are tentatively assigned as (5/2$^+$) ground state and (3/2$^+$) excited state. The first-excited 1/2$^+$ state which is predicted at energy of 0.3--0.4 MeV above the g.s.~was not identified yet, though its population may explain the 55-mrad peak in the experimental angular-correlation pattern in Fig.~\ref{fig:theta-p-Mg_b1}. The mass excess of the $^{21}$Al g.s.~was derived from the measured $Q_{1p}$ value to be +25.9(1) MeV, which is a challenging test for the predictions by nuclear mass models.

The observation of unbound $^{21}$Al g.s.~leads to the following statements about its neighboring nuclei in the nuclear chart. (1) Together with the very recent mass measurement of $^{22}$Al which provides its 1\textit{p}-separation energy of about 100 keV~\cite{Sun:2024,Campbell:2024}, one may conclude that the proton dripline is established in the Al isotopic chain.~(2) The loosely bound isotope $^{22}$Si (recent mass measurement for this nucleus proves that it is bound~\cite{Zhang:2024}) is a two-proton Borromean nucleus whose three-body configuration $^{20}$Mg+\textit{p}+\textit{p} has no bound sub-system $^{20}$Mg+\textit{p}$\rightarrow^{21}$Al.

{\textbf{Acknowledgments.}} 
This work was supported in part by the Helmholtz International Center for FAIR (HIC for FAIR); the Helmholtz Association (grant IK-RU-002); the European Union’s Horizon Europe Research and Innovation programme under Grant Agreement No 101057511 (EURO-LABS); the Chinese Academy of Sciences President's International Fellowship Initiative (Grant No. 2024PVA0005); the Russian Science Foundation (Grant No. 22-12-00054); the Polish National Science Center (Contract No. 2019/33/B/ST2/02908); the Helmholtz-CAS Joint Research Group (grant HCJRG-108); the Ministry of Education \& Science, Spain (Contract No. FPA2016-77689-C2-1-R); the Ministry of Economy, Spain (grant FPA2015-69640-C2-2-P); the Hessian Ministry for Science and Art (HMWK) through the LOEWE funding scheme; the Justus-Liebig-Universit\"at Giessen (JLU) and the GSI under the JLU-GSI strategic Helmholtz partnership agreement; and DGAPA-PAPIIT IG101423. This work was carried out in the framework of the Super-FRS Experiment Collaboration.

\bibliographystyle{apsrev4-2}

\end{document}